\documentstyle[preprint,aps,prd]{revtex}

\begin{document}

\title {Anomalies and symmetries of the regularized action}

\author{C\'esar~D.~Fosco$^{1}$~\footnote{Electronic address:
    fosco@cab.cnea.gov.ar} and Francisco D.\ 
  Mazzitelli$^{2}$~\footnote{Electronic address: fmazzi@df.uba.ar}}

\address{{\it $^1$ Centro At\' omico Bariloche\\ 8400 Bariloche, Argentina\\ 
    $^2$ Departamento de F\'\i sica, Facultad de Ciencias Exactas y
    Naturales\\ Universidad de Buenos Aires - Ciudad Universitaria, Pabell\'
    on I\\ 1428 Buenos Aires, Argentina}}

\maketitle

\begin{abstract}
  We show that the Pauli-Villars regularized action for a scalar field
  in a gravitational background in $1+1$ dimensions has, for any value
  of   the   cutoff  $M$,   a   symmetry   which  involves   non-local
  transformations   of   the  regulator   field   plus  (local)   Weyl
  transformations  of the  metric tensor.   These  transformations, an
  extension  to the  regularized  action of  the  usual Weyl  symmetry
  transformations   of   the  classical   action,   lead   to  a   new
  interpretation   of  the   conformal   anomaly  in   terms  of   the
  (non-anomalous) Jacobian for  this symmetry.  Moreover, the Jacobian
  is automatically regularized, and yields the correct result when the
  masses  of the  regulators  tend  to infinity.   In  this limit  the
  transformations, which  are non-local  on a scale  of $\frac{1}{M}$,
  become  the  usual Weyl  transformations  of  the  metric.  We  also
  present the example of the chiral anomaly in $1+1$ dimensions.
\end{abstract}

\newpage

\section{Introduction}

Anomalies are one of the most stricking manifestations of the presence
of ultraviolet infinities in quantum field theory phenomena. They have
important  consequences  for the  quantum  consistency (unitarity  and
renormalizability) of  different models, as  well as for the  study of
the   theoretical  structure  of   quantum  field   theory,  providing
interesting relations with mathematical objects.

The  regularization of  a  theory is  a  procedure with,  by its  very
definition,  produces a  violent  modification in  its large  momentum
behaviour.  Anomalies  arise  when  this  modification,  whatever  the
particular  regularization   method  applied,  violates   a  classical
symmetry.  In the  resulting  renormalized quantum  field theory,  the
quantum   counterpart  of  that   symmetry  becomes   `anomalous'.  In
particular,  continuum   symmetries  will   no  longer  lead   to  the
conservation of a current, thus modifying the naive Ward identities.

In  the  functional  integral  representation, anomalies  are  usually
attributed to the  fact that, while the classical  action is invariant
under a given symmetry transformation, the integration measure is not.
When  performing a  change of  variables associated  to  the classical
symmetry, the  Jacobian of the  transformation is, due  to ultraviolet
divergences,   not  well   defined.  When   carefully   evaluated,  by
introducing a  proper regularization, this  Jacobian gives rise  to an
anomalous term in  the effective action~\cite{Fujikawa}, or, depending
on the context, to anomalous Ward identities.

In this approach, the  symmetry transformations are implemented on the
{\em  unregularized\/} action,  which is  formally invariant,  and the
regularization  is  only implemented  afterwards,  when computing  the
Jacobian~\footnote{The  regularization of this  Jacobian is  of course
related  to  the  definition  of  the integration  measure.}   of  the
transformation. It  is worth  remarking that the  unregularized action
contains field  components at all  the momentum scales,  in particular
the ones that are above the cutoff.  To avoid this unpleasant feature,
we shall start by considering the regularized action, and restrict our
study  to transformations  that leave  this functional  invariant.  Of
course,   we  shall   impose   the  constraint   that  the   resulting
transformations  should tend  to the  usual  ones when  the cutoff  is
removed.  Also,  due to  the fact that  they avoid the  large momentum
modes,  they will necessarily  have to  be non  local, although  in a
scale of the order of the inverse of the cutoff.

As a particularly convenient form  of the regularized action, which we
will  use in  this note,  we shall  introduce a  set  of Pauli-Villars
regulator  fields, with  carefully tunned  masses $M_i$  that  tend to
infinity at  the end of the  calculation.  Due to the  presence of the
regulator masses,  the regularized action will be  no longer invariant
under some  the classical symmetries,  namely, those that rely  on the
masslessness of the fields in the action.

In the present letter, we  will point out that this regularized action
{\it does\/}  have a generalized, non local  symmetry, that reproduces
the classical symmetry in the limit $M_i\rightarrow \infty$. Moreover,
when performing  a change of variables  in the path  integral based on
this symmetry,  the associated  Jacobian is {\em  finite}, reproducing
the anomalous term  in the effective action, when  the cutoff tends to
infinity.

In section~\ref{sec:conf} we discuss  this procedure in detail for the
gravitational  conformal anomaly of  a massless  real scalar  field in
$1+1$  dimensions.  We  exhibit the  explicit  form of  the non  local
symmetry of the regularized  action and compute the conformal anomaly,
which  of course  agrees with  the  Liouville action~\cite{Polyakov1},
\cite{Polyakov2}.   In section~\ref{sec:chiral}  we  consider massless
fermions   in  $1+1$   dimensions  coupled   to   the  electromagnetic
field.  Again  we  present   a  generalized  chiral  symmetry  of  the
regularized    action     and    compute    the     chiral    anomaly.
Section~\ref{sec:concl} contains our conclusions.

\section{The conformal anomaly in $1+1$ dimensions}\label{sec:conf}  
In this  section we shall  consider the partition  function ${\mathcal
  Z}[g_{\mu\nu}]$,  corresponding to  a massless  scalar field  in the
  presence   of  an   external  gravitational   background   in  $1+1$
  dimensions:
\begin{equation}\label{eq:defzg}
  {\mathcal  Z}[g_{\mu\nu}]\;=\;  \int [{\mathcal  D}  \varphi 
]_g \;  e^{i
    S[\varphi \,,\, g_{\mu\nu}]}
\end{equation}
where
\begin{equation}\label{eq:defs}
 S[\varphi ,  g_{\mu\nu}]\;=\; \frac{1}{2} \int d^2 x  \, \sqrt{-g} \,
g^{\mu\nu} \partial_\mu \varphi \partial_\nu \varphi
\end{equation}
and \mbox{$g = \det  (g_{\mu\nu})$}. 
In Eq. (\ref{eq:defzg}) we have made it explicit the fact that
the definition of the scalar field integration measure
depends upon the background metric $g_{\mu\nu}$.

The classical action $S[\varphi ,
g_{\mu\nu}]$ is invariant under Weyl transformations of the metric:
\begin{equation}\label{eq:weyl}
g_{\mu\nu} \, \to g^\omega_{\mu\nu}(x) = \omega (x) g_{\mu\nu}(x)
\end{equation} 
\begin{equation}\label{eq:class}
S[\varphi , g^\omega_{\mu\nu}] \,=\, S[\varphi , g_{\mu\nu}]
\end{equation}
for any (strictly positive) $\omega (x)$. In adequate coordinates, any
metric in two dimensions is conformally flat:
\begin{equation}\label{eq:flat}
g_{\mu\nu} \,=\, e^{\sigma(x)} \eta_{\mu\nu}
\end{equation}
where $\eta_{\mu\nu}  = {\rm diag} (1,-1)$ denotes  the flat Minkowski
metric.  The classical  dynamics of a scalar field  in a gravitational
background in $1+1$ dimensions is therefore trivial:
\begin{equation}\label{eq:class1}
S[\varphi  ,  g_{\mu\nu}]   \,=\,  S[\varphi  ,  \eta_{\mu\nu}]  \,=\,
 \frac{1}{2}   \int  d^2   x   \,  \eta^{\mu\nu}\partial_\mu   \varphi
 \partial_\nu \varphi \;.
\end{equation}
This  conclusion  does not  hold  true,  of  course, for  the  quantum
dynamics, the  reason being  the existence of  the conformal  or trace
anomaly,   which  spoils  the   symmetry  under   the  transformations
(\ref{eq:weyl})  and produces  a  non vanishing  trace  in the  energy
momentum tensor~\cite{trace}.

In the usual setting, one  derives the quantum effects by dealing with
the vacuum functional (\ref{eq:defzg}).
Under  a Weyl transformation  (\ref{eq:weyl}), and using
the property (\ref{eq:class}), we see that
\begin{equation}\label{eq:usual2}
 {\mathcal     Z}[g_{\mu\nu}^\omega]\;=\;    \int     [{\mathcal    D}
  \varphi]_{g^\omega} \; e^{i S[\varphi , g_{\mu\nu}]}\;.
\end{equation}
Namely, any possible quantum  effect must come from the non-invariance
of the integration measure under Weyl transformations of the metric.

As  shown by  Polyakov~\cite{Polyakov1} -  \cite{Polyakov2}, there  is,
indeed, an anomalous Jacobian, the exponential of the Liouville action
\begin{equation}\label{eq:usual3}
[{\mathcal  D} \varphi]_{g^\omega}\,=\,  [{\mathcal  D} \varphi]_g  \,
J[\sigma]\, =\, [{\mathcal  D} \varphi]_g \; \exp{i {\cal A}[g_{\mu\nu},w]}
\end{equation}
where ${\cal  A}[g_{\mu\nu},w]$ is the anomaly, a  local functional of
$g_{\mu\nu}$.

By using the decomposition (\ref{eq:flat}), and normalizing ${\mathcal
Z}$ such that ${\mathcal Z}[\eta_{\mu\nu}] = 1$, one also sees that
\begin{equation}\label{eq:decp}
 {\mathcal   Z}[g_{\mu\nu}]\;=\;   J[\sigma]\;=\;   \exp{i   {\mathcal
 A}[w]}\;,
\end{equation}
where ${\mathcal A}[w] \equiv {\mathcal A}[\eta_{\mu\nu}, w]$ is given
by
\begin{equation}\label{Polyakov}
{\cal     A}[w]    =     {1\over    96\pi}\int     d^2x{1\over    w^2}
\partial_{\mu}w\partial^{\mu}w \;.
\end{equation}
As  can be  easily  proved  from the  perturbative  derivation of  the
anomaly given  in Ref. \cite{Polyakov2},  when the system  consists of
$d$ bosonic  scalar fields  and $\bar d$  Grassmann scalar  fields the
Jacobian becomes $J[\sigma]\;=\; \exp{i(d-\bar d) {\cal A}[g,w]}$.

In what follows  we will show that, if one  works with the regularized
action rather  than the classical  action $S[\varphi ,g]$, there  is a
non local  symmetry, and one can  compute the anomaly  as the Jacobian
associated   to  a  non   local  redefinition   of  the   fields.  The
Pauli-Villars method, first applied  to this system in \cite{bernard},
amounts   to   the    introduction   of   three   `regulator   fields'
\mbox{${\bar\chi},\chi,\eta$} and a cutoff $M$ so that
$$
S^{reg}[\varphi,  {\bar\chi},\chi,\eta,g_{\mu\nu},M] \;=\; \frac{1}{2}
\int  d^2  x \,  \sqrt{-g}  \,  \left( g^{\mu\nu}  \partial_\mu\varphi
\partial_\nu\varphi      +      g^{\mu\nu}      \partial_\mu{\bar\chi}
\partial_\nu\chi - M^2 {\bar\chi}\chi \right.
$$
\begin{equation}\label{eq:defsreg} 
\left. +  g^{\mu\nu} \partial_\mu\eta \partial_\nu\eta -  2 M^2 \eta^2
\right)
\end{equation} 
where ${\bar\chi},\chi$ are complex  Grassmann fields, while $\eta$ is
a real scalar field. The number and massess of the regulators are just
right  to   make  the  vacuum  energy  in   a  non-trivial  background
finite. For the sake  of convenience, we rewrite (\ref{eq:defsreg}) in
a more compact form as follows:
\begin{equation}\label{eq:sreg} 
S^{reg}[\varphi,{\bar\chi},\chi,\eta,g_{\mu\nu},M]  \;=\; -\frac{1}{2}
\left[\langle       \varphi      |{\Box'}|\varphi\rangle      +\langle
{\bar\chi}|({\Box'}+M^2\sqrt{-g})|\chi\rangle +\langle\eta |({\Box'}+2
M^2\sqrt{-g})|\eta\rangle\right]
\end{equation} 
where
\begin{equation}\label{eq:defbt} 
{\Box'} \;=\;  \sqrt{-g} \Box \;=\;  \partial_\mu \sqrt{-g} g^{\mu\nu}
\partial_\nu
\end{equation} 
and $\Box$ is the usual curved space Laplacian operator
$\Box   \;=\;   \frac{1}{\sqrt{-g}}\partial_\mu  \sqrt{-g}   g^{\mu\nu}
\partial_\nu$.
We used  a Dirac braket like  notation for the scalar  product in {\em
flat\/} two dimensional spacetime:
\begin{equation}\label{eq:defprd} 
\langle \phi_1 | \phi_2  \rangle \,=\, \int d^2x \,\phi_1(x) \phi_2(x)
\;,
\end{equation} 
which is convenient, since we  have absorbed the factor $\sqrt{-g}$ of
the measure in  the operators. It is worth  noting that both ${\Box'}$
and  $\sqrt{-g}$  are symmetric  (real  Hermitian)  operators for  the
scalar  product  (\ref{eq:defprd}).   Also,  ${\Box'}$  is  explicitly
invariant under Weyl transformations.

Due to the presence of the  mass terms for the regulator fields, it is
evident  that  the regularized  action  is  not  invariant under  Weyl
transformations of the metric tensor:
$$
S^{reg}[\varphi,    {\bar\chi},\chi,\eta,g^\omega_{\mu\nu},M]    =
-\frac{1}{2}  \left[\langle  \varphi |{\Box'}|\varphi\rangle  +\langle
{\bar\chi}|({\Box'}+M^2\omega    \sqrt{-g})|\chi\rangle   +\langle\eta
|({\Box'}+2 M^2 \omega \sqrt{-g})|\eta\rangle\right]
$$
\begin{equation}\label{eq:sreg1}
\neq\; S^{reg}[\varphi, {\bar\chi},\chi,\eta,g_{\mu\nu},M] \;.
\end{equation}
We  may, however,  compensate the  non  invariance of  $S^{reg}$ by  a
transformation of the regulator fields:
\begin{eqnarray}
\vert\varphi^w> &=& \vert\varphi >\nonumber\\ \vert\eta^w> &=& (\Box'+
2   M^2  w\sqrt{-g})^{-1/2}  (\Box'+   2  M^2\sqrt{-g})^{1/2}\vert\eta
>\nonumber\\ \vert\chi^w>  &=& (\Box'+ M^2  w\sqrt{-g})^{-1/2} (\Box'+
M^2\sqrt{-g})^{1/2}\vert\chi >
\label{nonlocal}
\end{eqnarray}
after which we see that
\begin{equation}
S_{reg}[\varphi^w,\eta^w,\bar\chi^w,\chi^w,       g^w,       M]      =
S_{reg}[\varphi,\eta,\bar\chi,\chi, g, M] \;.
\label{symmetry}
\end{equation}
Eqs. (\ref{nonlocal} -\ref{symmetry}) define the non local symmetry of
the regularized action. This symmetry must be studied of course at the
quantum level, by considering the regularized vacuum functional
\begin{equation}\label{eq:defzreg}
Z_{reg}[g_{\mu\nu}] \;=\;  \int[{\cal D}\varphi {\cal  D}\eta {\cal D}
\bar\chi  {\cal D}\chi]_g  \exp(i S_{reg}[\varphi,\eta,{\bar\chi},\chi
,g,M]) \;.
\end{equation}
Performing the Weyl transformation (\ref{eq:weyl}) for $g_{\mu\nu}$ in
(\ref{eq:defzreg})    followed   by    the    change   of    variables
(\ref{nonlocal}) in the functional integral, we see that:
\begin{eqnarray}
Z_{reg}[g_{\mu\nu}^w] &=& \int[{\cal D}\varphi^w{\cal D}\eta^w{\cal D}
\bar\chi^w{\cal                 D}\chi^w]_{g^w}                 \exp(i
S_{reg}[\varphi,\eta,{\bar\chi},\chi,g,M])\nonumber\\               &=&
J_{reg}[g,w,M] \int  [{\cal D}\varphi  {\cal D}\eta {\cal  D} \bar\chi
{\cal D}\chi]_g \exp(i S_{reg}[\varphi,\eta,{\bar\chi},\chi ,g,M])
\label{zreg}
\end{eqnarray}
where   $J_{reg}$  denotes   the  Jacobian   for   the  transformation
(\ref{nonlocal}):
$$
J_{reg}[g,w,M] \;=\; \det [({\Box'} + M^2\sqrt{-g})^{-1} (\Box'+ M^2 w
\sqrt{-g})]
$$
\begin{equation}
\times  \det  [({\Box'}  +  2  M^2  \sqrt{-g})^{-1}  ({\Box'}+2  M^2  w
\sqrt{-g})]^{-1/2} \;,
\label{jreg}
\end{equation}
and  the suffix  `reg'  is used  because,  as we  will  see now,  this
Jacobian is finite.  It is worth noting that, although the integration
measure for  each field $[{\cal  D} \phi_i]$ depends non  trivially on
the background  metric, as in  the unregularized case, the  product of
the integration measures  for the four fields does  {\em not\/} depend
on  $w$,  due  to  the  cancellation  between  the  anomalous  factors
corresponding to bosonic and Grassmann fields.  These Jacobian factors
are independent of the masses of the fields.

If  we define  the  finite quantities  $Z$  and $J$  as  the limit  of
$Z_{reg}$ and $J_{reg}$ for $M\rightarrow\infty$ equation (\ref{zreg})
implies that $ Z[g^w] \;=\; J[g,w]\, Z[g] $.  Thus we have to evaluate
Eq.(\ref{jreg}) in the limit $M\rightarrow\infty$.

Let  us now  calculate the  regulated Jacobian  (\ref{jreg}).   As the
metric is assumed  to be conformally flat, and  ${\Box'}$ is invariant
under (\ref{eq:weyl}), it is obvious  that we may replace ${\Box'}$ by
$\Box$,  and that  $\sqrt{-g}=1$.  We  then rewrite  $J_{reg}$  in the
form:
\begin{equation}\label{eq:jreg1}
 J_{reg}[\eta,w,M]\;=\;          J^{(1)}_{reg}[w,M]         \;\times\;
 J^{(2)}_{reg}[w,M]
\end{equation}
where
\begin{eqnarray}
 J^{(1)}_{reg}[w,M]  &=&   \det  \left[(\Box  +   M^2)^{-1}(\Box+  M^2
w)\right]  \nonumber\\ J^{(2)}_{reg}[w,M]  &=&  \left[ J^{(1)}_{reg}[
w,\sqrt{2}M] \right]^{-1/2} \;.
\label{eq:jreg2}
\end{eqnarray}
We  shall now,  for calculational  purposes,  consider $J^{(1)}_{reg}$
alone, since  the factor  $J^{(2)}_{reg}$ can be  obtained from  it by
some simple substitutions.  However, the  two factors have to be taken
together for the cancellation between UV divergences to happen.

To evaluate  $J^{(1)}_{reg}[w,M]$, we take into account  the fact that
we will, in  the end, be interested in the $M  \to \infty$ limit. This
justifies  the   use  of   some  form  of   expansion  in   powers  of
$\frac{1}{M}$. A  small dimensionless parameter  has then to  be built
using  $M$, and  the only  other dimensionful  object:  derivatives of
$\omega$  ($\omega$  itself  is  dimensionless). We  then  follow  the
derivative  expansion  technique~\cite{aitchi}   to  split  the  field
$\omega$  into   a  slowly   varying  part  ${\tilde\omega}$   plus  a
fluctuating piece $\alpha$
\begin{equation}\label{eq:split}
  \omega(x) \;=\; {\tilde\omega}(x) + \alpha (x) \;,
\end{equation}
where ${\tilde\omega}(x)$ is  to be regarded as a  constant when acted
by the derivative operator. We then rotate to Euclidean spacetime, and
expand $\ln J^{(1)}_{reg}[w,M]$ in powers of $\alpha$, starting from
\begin{equation}\label{eq:exp1}
\ln  J^{(1)}_{reg}[w,M]\;=\; {\rm  Tr} \ln  ( \frac{-\partial^2  + M^2
{\tilde\omega}}{-\partial^2  +  M^2}  )  +  {\rm  Tr}  \ln  \left(1  +
\frac{M^2 \alpha}{-\partial^2 + M^2 {\tilde\omega}} \right)\;,
\end{equation}
where  $\partial^2\,=\,\delta_{\mu\nu}\partial_\mu\partial_\nu$ is the
flat, Euclidean spacetime Laplace  operator.  Taking into account that
the linear term  vanishes, and terms with more  than two $\alpha$'s are
suppressed  by negative powers  of $M$,  it is  sufficient to  use the
expansion:
$$
\ln  J^{(1)}_{reg}[w,M]\;=\; {\rm  Tr} \ln  ( \frac{-\partial^2  + M^2
  {\tilde\omega}}{-\partial^2 + M^2} )
$$
\begin{equation}\label{eq:exp2}
-\frac{1}{2} {\rm Tr} \left(\frac{M^2 \alpha }{-\partial^2 + 
M^2{\tilde\omega}} \; 
\frac{M^2 \alpha }{-\partial^2 + M^2{\tilde\omega}} \right) \;+\;
{\mathcal O}(\frac{1}{M^2})\;.
\end{equation}
The  first,  zero  derivative  term,  is divergent  (even  for  finite
$M$). However,  remembering that we are using  a Pauli-Villars scheme,
we have to evaluate the  momentum integral that results from combining
it  with  the  corresponding  contribution from  $\ln  J^{(2)}$.  This
produces a finite answer:
\begin{equation}
{\rm  Tr}\ln (\frac{-\partial^2  +  M^2 {\tilde\omega}}{-\partial^2  +
M^2}  )  \;-\;\frac{1}{2}  {\rm  Tr}\ln (\frac{-\partial^2  + 2 M^2  {
\tilde\omega}}{-\partial^2 + 2  M^2} )\; =\; \frac{M^2 \ln  2 }{4 \pi}\,
({\tilde \omega}-1) \;,
\label{eq:w-1}
\end{equation}  
proportional  to $M^2$.   This is  finite (for  finite $M$),  and this
shows that the Jacobian for  the non local symmetry transformations is
indeed  finite.    Of  course,  when  $M$  tends   to  infinite,  this
contribution   diverges. The term proportional to $\tilde\omega$
requires   the   introduction  of   a
counterterm of the cosmological constant type. 
The $\tilde\omega$-independent
divergence in Eq. \ref{eq:w-1} can be 
absorbed into the normalization factor of
$\mathcal Z$.

For the  second order term  (which is finite  when $M \to  \infty$), a
standard calculation yields, for $M\to\infty$
\begin{equation}\label{eq:exp3}
-\frac{1}{2}       {\rm       Tr}      [\frac{M^2}{-\partial^2       +
M^2{\tilde\omega}}\alpha            \frac{M^2}{-\partial^2           +
M^2{\tilde\omega}}\alpha ]  = -{1\over 48\pi}\int  d^2x{1\over {\tilde
w}^2}\partial_{\mu} \alpha\partial^{\mu}\alpha \;.
\end{equation}
The derivative expansion technique implies~\cite{aitchi}, on the other
hand, that ${\tilde  \omega}$ may be replaced by  $\omega$ in a second
order  term,  and  that  derivatives  of $\alpha$  are  tantamount  to
derivatives of $\omega$. Then,
\begin{equation}\label{eq:exp4}
 -\frac{1}{2}       {\rm      Tr}       [\frac{M^2}{-\partial^2      +
M^2{\tilde\omega}}\alpha            \frac{M^2}{-\partial^2           +
M^2{\tilde\omega}}\alpha   ]   =   -{1\over   48\pi}\int   d^2x{1\over
\omega^2}\partial_{\mu}\omega\partial^{\mu}\omega \;.
\end{equation}
This contribution has to be combined with the second order term coming
from $J_{reg}^{(2)}$,  which only  differs in a  $-\frac{1}{2}$ global
factor.  Then,
\begin{equation}\label{eq:exp5}
\lim_{M\to\infty}     J_{reg}[\eta,w,M]\,=\,    -{1\over    96\pi}\int
d^2x{1\over \omega^2} \partial_{\mu}\omega\partial^{\mu}\omega \;.
\end{equation}
which is the (Euclidean) Liouville action.  Rotating back to Minkowski
spacetime we obtain the result given in Eq.  (\ref{eq:decp}).

\section{The chiral anomaly}\label{sec:chiral}
This  example shares  many properties  with  the previous  one of  the
conformal anomaly, and  helps to understand the general  nature of the
procedure we have applied in section~\ref{sec:conf}.

We shall consider here ${\mathcal  Z}[A]$, the vacuum functional for a
massless fermion in $1+1$ dimensions,
\begin{equation}\label{eq:defza}
{\mathcal Z}[A] \;=\; \int  [{\mathcal D} {\bar\psi} {\mathcal D} \psi
]_A \; e^{i S_F[{\bar\psi},\psi;A]}
\end{equation}
with
\begin{equation}\label{eq:defsf}
  S_F[{\bar\psi},\psi;A]\;=\;  \int  d^2x  \,  {\bar\psi} (i  \not  \!
  \partial - e \not \!\! A ) \psi \;.
\end{equation}
Again,  the  integration  measure  depends  on  the  background  field
configuration.  The  background gauge field  $A$ may be  decomposed as
follows:
\begin{equation}\label{eq:decompos}
e A_\mu\;=\; \partial_\mu \varphi + \epsilon_{\mu\nu}\partial_\nu \rho
\;,
\end{equation}
where  $\varphi$  and  $\rho$  are  scalar  and  pseudoscalar  fields,
respectively.  Then one  easily sees that the fermionic  action may be
rewritten as
\begin{equation}\label{eq:sf1}
  S_F[{\bar\psi},\psi;A]\;=\;  \int  d^2x  \,  {\bar\psi} (i  \not  \!
  \partial - \not \! \partial(\varphi + \gamma_5 \rho) \psi \;,
\end{equation}
what means that the gauge field may actually be erased by a gauge plus
chiral  transformation of  the  fermions. This,  as  for the  previous
example, implies that the classical dynamics of the system is trivial.
However, the anomaly under chiral fermionic transformations introduces
a non trivial quantum dynamics.

In the  usual derivation, one  performs chiral transformations  of the
fermions,
\begin{equation}\label{eq:chrt}
  \psi  (x) \;=\;  e^{i \alpha  (x) \gamma_5}  \psi  (x) \;\;\;,\;\;\;
  {\bar\psi} (x) \;=\; {\bar\psi} (x) e^{i \alpha (x) \gamma_5}\;,
\end{equation}
and  the  chiral  anomaly  appears  from  the  non-invariance  of  the
fermionic measure.

To consider an alternative  derivation, we note that the Pauli-Villars
regularized  action in  this case  requires the  addition of  just one
(massive) bosonic spinor field $\phi$, such that
\begin{equation}\label{eq:defsfr}
  S^{reg}_F[{\bar\psi},\psi,{\bar\phi},\phi;A]\;=\;   \int   d^2x   \,
\left[  {\bar\psi} (i \not  \!  \partial  - e  \not \!\!   A )  \psi +
{\bar\phi} (i \not \!\partial - e \not \!\! A - M) \phi \right]\;.
\end{equation}

When  $\alpha$  is  a  constant, the  non-local  infinitesimal  chiral
symmetry transformations of (\ref{eq:defsfr}) are,
\begin{eqnarray}
\delta  \psi  &=&  i   \,\alpha  \,\gamma_5  \psi  \nonumber\\  \delta
{\bar\psi}  &=& i  \, \alpha  \,{\bar\psi}\gamma_5  \nonumber\\ \delta
\phi &=& i  \, \alpha \, \gamma_5 \frac{\not \!\! D}{\not  \!\!  D - i
M}\phi   \nonumber\\   \delta   {\bar\phi}   &=&  i   \,   \alpha   \,
{\bar\phi}\frac{\not \!\! D}{\not \!\! D - i M} \gamma_5 \;,
\label{eq:regftr}
\end{eqnarray}
where
\begin{equation}\label{eq:defd}
 \not \!\! D \;=\; \not \!\partial + i e \not \!\! A \;.
\end{equation}

The  action  is  invariant  under  these  transformations,  while  the
Jacobian becomes
\begin{equation}\label{eq:chirj}
  J =  \det\left[ 1 + i \alpha  \gamma_5 (1 - \frac{\not  \!\! D}{\not \!\!
    D-i M}) \right]^{-2}
\end{equation}
which may be rewritten as
\begin{equation}\label{eq:chirj1}
J = \exp \left[ -2 i \alpha {\rm Tr} \gamma_5 ( \frac{1}{1 + \frac{\not \! 
D^2}{M^2}}) \right]\;.
\end{equation}
The funtional trace  is finite, and reproduces the  proper result when
$M \to \infty$:
\begin{equation}\label{eq:chres}
J = \exp[ i \frac{e}{2 \pi} \alpha \int d^2x \epsilon^{\mu\nu}\partial_\mu
A_\nu]\;.
\end{equation}
When  $\alpha$ is spacetime  dependent, the  regularized action  is no
longer invariant. However,  we may use that kind  of transformation to
get  rid of  the  dependence in  $A_\mu$.   Those transformations  are
defined by
$$
\delta  \psi  \;=\; i  \,\alpha(x)  \,\gamma_5  \psi \;\;,\;\;  
\delta {\bar\psi}  \;=\; i  \,{\bar\psi}\gamma_5  \alpha(x)  
$$
\begin{equation}
\delta \phi \;=\; i \, \gamma_5 
\frac{\not \!\! D}{\not  \!\!  D - i M} \alpha(x) \phi    \;\;,\;\; 
\delta {\bar\phi} \;=\; i \,{\bar\phi}\alpha(x) 
\frac{\not \!\! D}{\not \!\! D - i M} \gamma_5 \;,
\label{eq:reglftr}
\end{equation}
and the corresponding variation of the action is
\begin{equation}\label{eq:delsfr}
  \delta S^{reg}_F \;=\;  -  \int  d^2x  \,
[  {\bar\psi}  \not\!\partial ( \gamma_5 \alpha )  \psi  +
{\bar\phi}\not\!\partial( \gamma_5\alpha) \phi ]\;.
\end{equation}
The Jacobian is easily shown to be
\begin{equation}\label{eq:chres1}
J = \exp[ i \frac{e}{2 \pi} \int d^2x  \alpha(x) \epsilon^{\mu\nu}\partial_\mu
A_\nu]\;.
\end{equation}

\section{Conclusions}\label{sec:concl}
We  have  presented  two   concrete  examples  of  systems  where  the
regularized  action has  a non  local  symmetry which  is the  natural
extension  of  the  standard  symmetry of  the  unregularized  action.
Moreover, the  application of those transformations  in the functional
integral  framework   yields  regularized  Jacobians   which  properly
reproduce the anomalies when the  cutoff tends to infinity. This makes
the connection between the regularization  of the diagrams of a model,
and the  regularization of its  Jacobian more transparent than  in the
usual setting.

We  remark that  our method  differs  also from  performing the  usual
(local)  transformations  to the  regularized  action. This  procedure
would give  {\em no  Jacobian}, due to  the cancellation  between bare
fields  and   regulators,  while  the  regularized   action  would  be
non-invariant.  The  anomaly  would  appear  in  this  case  from  the
non-invariance of the regularized action.

Finally,  we want  to remark  that the  phenomenon we  have described,
namely, the existence  of a remnant of the  classical symmetry for the
regularized   action    is   not   new.    It    has   been   recently
emphasized~\cite{luscher} that massless  fermions on the lattice, even
thought the regulatization breaks  the naive chiral symmetry, may have
a lattice equivalence of that  symmetry, if the lattice Dirac operator
satisfies  the  Ginsparg-Wilson  relation~\cite{gins}.   Indeed,  this
relation  can be used  to derive  the chiral  anomaly and  the related
index theorems on the lattice~\cite{Fujikawa1}.
 
\acknowledgments This work was supported by Universidad de Buenos Aires,
CONICET (Argentina), Fundacion Antorchas, ANPCyT, CNEA and the Abdus Salam
International Centre for Theoretical Physics.

The authors would  like to thank the Abdus  Salam International Centre
for  Theoretical Physics  for  hospitality during  completion of  this
work.

\end{document}